\documentclass[twocolumn,showpacs,preprintnumbers,amsmath,amssymb,superscriptaddress]{revtex4}

\usepackage{longtable}
\usepackage{graphicx}
\usepackage{dcolumn}
\usepackage{bm}

\begin{document}

\title{Impact of Low-Energy Response to Nuclear Recoils in Dark Matter Detectors}

\newcommand{\usd}{Department of Physics, The University of South Dakota, Vermillion, South Dakota 57069}
\newcommand{\yru}{College of Physics, Yangtz River University, Jingzhou 434023, China}
\newcommand{\dmm}{Dongming.Mei@usd.edu}
\newcommand{\wzw}{Wenzhao.Wei@usd.edu}
\newcommand{\lw}{Lu.Wang@usd.edu}

\author{     D.-M.~Mei          }\altaffiliation[Corresponding Author: ]{\dmm}	\affiliation{ \usd } 	\affiliation{ \yru }
\author{W.-Z. Wei} \altaffiliation[Equal contributor:] {\wzw} \affiliation{ \usd} 
\author{L. Wang}\altaffiliation[Equal contributor:] {\lw} \affiliation{ \usd}
 			

\begin{abstract}
We report an absolute energy response function to electronic
and nuclear recoils for germanium and liquid xenon detectors.
As a result, we show that the detection energy threshold of nuclear recoils for 
a dual-phase xenon detector can be a few keV for a given number of detectable photoelectrons.  
We evaluate the average energy expended per electron-hole pair to be $\sim$3.32 eV,
which sets a detection energy threshold of $\sim$2.15 keV for a germanium detector at 50 mini-Kelvin at 69 volts
with a primary phonon frequency of 1 THz.
The Fano factors of nuclear and electronic recoils that constrain the capability for discriminating nuclear 
recoils below 2-3 keV recoil energy for both
technologies are different.  
\end{abstract}

\pacs{95.35.+d, 95.30.Cq, 07.05.Tp, 29.40.Wk}

\maketitle
Observations from the 1930’s~\cite{zwicky} led to the idea of dark matter, more recent observations~\cite{hlk, rjg} lead to 
the contemporary and shocking revelation that 96\% of the matter 
and energy 
in the universe neither emits nor absorbs electromagnetic radiation. A popular candidate for dark matter is 
the WIMP (Weakly Interacting Massive Particle), a particle with mass thought to be comparable to heavy nuclei, 
but with a feeble and extremely short-range interaction with atomic nuclei. Theories involving Supersymmetry (SUSY), 
which are being tested at the Large Hadron Collider~\cite{lhc}, naturally imply a particle that could be the WIMP~\cite{jlf}. 
While WIMPs appear to collide with atomic nuclei very rarely, they would cause atoms to recoil at a velocity 
several thousand times the speed of sound when they do interact~\cite{mwg}. 
Direct detection of WIMPs requires the ability to observe low-energy recoils and discriminate electronic recoils 
originating from background events from nuclear recoils induced by WIMPs.
   
The detector response to low-energy recoils can be described as $E_{eff} = Y(E_{r}) \times E_{r}$,
where $Y(E_{r})$ is the so called ionization efficiency. For electronic recoils, $Y(E_{r})$ = 1, while for 
nuclear recoils, $Y(E_{r})$ is a function of the nuclear recoil energy~\cite{lind}. For a given experiment, the total visible energy can be 
expressed as
\begin{equation}
E_{vis} = \frac{\alpha}{1+\beta \frac{dE_{eff}}{dx}} E_{eff},
\label{eq:mei1}
\end{equation}
where $\alpha$ and $\beta$ can be determined both theoretically or experimentally using known energy sources while $\frac{dE_{eff}}{dx}$ is
given by a theoretical calculation~\cite{mei}. Eq.(\ref{eq:mei1}) is in a form of Birks' law~\cite{birks} that takes into account
the reduction of visible energy due to the recombination of electron-hole (e-h) pairs or the reduction of scintillation yield 
along a high-density ionization track. This is to say that the reduction of visible energy is only a function of ionization 
density regardless of particle type. 
For electronic recoils, in the case in which the product of $\beta\frac{dE_{r}}{dx}$ 
is insignificant, $E_{vis}$ is a linear function of $E_{r}$. Otherwise,
 $E_{vis}$ may not be a linear function of $E_{r}$ if the product of $\beta\frac{dE_{r}}{dx}$ is significant. In the
case of nuclear recoils, Eq.(\ref{eq:mei1}) becomes
\begin{equation}
E_{vis} = \frac{\alpha}{1+\beta \frac{dE_{eff}}{dx}}\times Y(E_{r}) \times E_{r}.
\label{eq:mei2}
\end{equation}
Because this depends on the product of $\beta\frac{dE_{eff}}{dx}$, the further reduction of visible energy in addition to
$Y(E_{r})$ needs to be taken into account. In the case of liquid xenon, $E_{vis}$ is evaluated by
many authors~\cite{mei, eap, lba, jth, mmi, tdo, ekp, wmu, man} including a data-driven model~\cite{ble, jmo, szy1, szy2}. 
Note that $E_{vis}$ is usually calibrated using gamma-ray sources 
and by assuming the
entire energy from well-known gamma-ray energies is detectable. This type of calibration is a relative calibration
because the real measurable energy for a given detector can be much smaller than the total energy deposition
from a given gamma energy. This is because a large fraction of deposited energy contributes to the creation 
of phonons due to the requirement of momentum conservation in the energy deposition process. This fraction of
energy cannot be measured by any generic ionization or scintillation detector, which can only measure the real energy to 
be the fraction of energy that creates e-h pairs and direct excitation, 
$E_{real} = \frac{E_{g}}{\varepsilon} \times E_{vis}$, 
 where $E_{g}$ is the band gap energy for a given target, and
$\varepsilon$ is the 
average energy required to produce an e-h pair. 

In a germanium ionization detector, the detectable charge is proportional to the number of e-h pairs,
 $N_{Q}$ = $\frac{E}{\varepsilon}$, created
by an incoming particle with energy deposition of $E$, where $\varepsilon$ = $E_{g}$ + 2$E_{f}$ + $rE_{R}$~\cite{fee} with $E_{g}$ the band gap energy,
$E_{f}$ the final retained energy less than $E_{g}$ for e-h pairs,  $r$ the number of phonons created per e-h pair, 
and $E_{R}$ the average energy per phonon. Note that $rE_{R}$ is temperature dependent.
  For a given temperature, $\varepsilon$ can be a constant down to very low energy~\cite{jmf}. Therefore, we
expect that the energy response function is an excellent linear function~\cite{wei}. 

In a dual-phase
xenon detector, $\varepsilon$ = (1.13 + 0.9$\frac{N_{ex}}{N_{i}}$+ 0.48)$E_{g}$~\cite{tta}, where $\frac{N_{ex}}{N_{i}}$ is the ratio of direct
excitation to ionization per energy deposition, which is energy dependent in which the direct excitation increases with energy decreasing. 
This implies that, in the low-energy region, the energy response function is not a linear function for a dual-phase xenon detector. Thus,
a relative calibration using well-known gamma-ray sources or electron sources above the low energy region 
can be largely uncertain in determining the energy scale for the low-energy region. This uncertainty can further
propagate in
the determination of the energy scale for nuclear recoils for a dual-phase xenon detector if a relative calibration is performed. 
Consequently,
the sensitivity limits achieved for WIMPs searches may possess large uncertainty.
We propose an absolute
energy response function (Eq.(\ref{eq:mei2})) to evaluate this uncertainty.

The determination of $\alpha$ and $\beta$ using the relationship 
between $E_{real}$ and $E_{r}$ is called an absolute energy response function. 
For a given detector, the variation of $\varepsilon$ results in a variation of real energy that can be detected. 
This variation can be predicted using the following relation:
\begin{equation}
\frac{E_g}{\varepsilon}\frac{dE_{vis}}{dx} = \frac{\alpha}{1+\beta \frac{dE_{eff}}{dx}} \frac{dE_{eff}}{dx}.
\label{eq:mei3}
\end{equation}

\begin{figure}
\includegraphics[angle=0., width=8.cm]{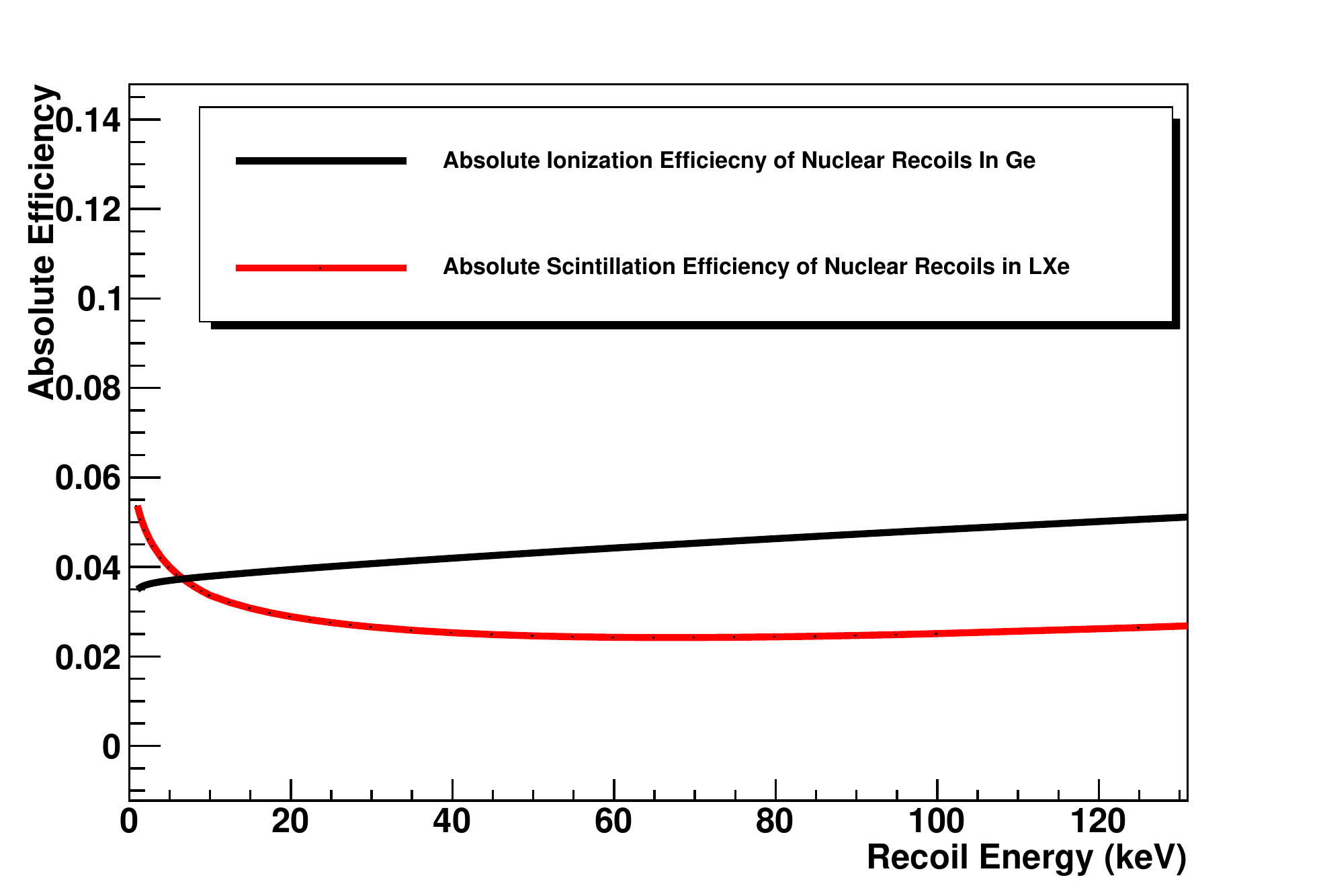}
\caption{\small {The absolute efficiency of nuclear recoils for liquid xenon at zero field and germanium detectors.}}
\label{fig:totaleff}
\end{figure}

 In the case of a germanium detector at 77 Kelvin, $\varepsilon$ = 2.96 eV and the variation of $\varepsilon$ is
negligible. Therefore, we predict $\alpha$ = (0.249$\pm$0.013) and $\beta$ = (5.12$\pm$2.68)$\times10^{-5}$ cm/MeV and 
these values are verified using experimental data~\cite{wei}. However, for a liquid xenon detector, the observable quanta 
are scintillation photons, which are produced through direct excitation and recombination of electron-ion (e-ion) pairs from
ionization. Thus, the average energy expended per e-ion pair varies as a function of the ratio of direct
excitation to ionization per energy deposition process, $\varepsilon$ = (1.13 + 0.9$\frac{N_{ex}}{N_{i}}$+ 0.48)$E_{g}$~\cite{tta}.
The ratio of direct excitation to ionization can be described as a function of deposition energy~\cite{sha}:
\begin{equation}
\frac{N_{ex}}{N_{i}} = \frac{1-exp(-\frac{I}{E})}{exp(-\frac{I}{E})},
\label{eq:mei4}
\end{equation}
where $I$ is the mean ionization potential and $E$ is the deposition energy. For xenon, $I$= 35.55 + 9.63$Z$~\cite{muk}, where $Z$=54.
As can be seen in Eq.(\ref{eq:mei4}), when $E$ = 1 keV, $\frac{N_{ex}}{N_{i}}$ = 0.74 
while $E$ = 10 keV, $\frac{N_{ex}}{N_{i}}$ = 0.057. Plugging Eq.(\ref{eq:mei4}) into
$\varepsilon$ = (1.13 + 0.9$\frac{N_{ex}}{N_{i}}$+ 0.48)$E_{g}$, one obtains $\varepsilon$. 
Using Eq.(\ref{eq:mei3}),
we determine $\alpha$ = (0.74$\pm$0.04) and $\beta$ = (6.57$\pm$1.1)$\times10^{-3}$ cm/MeV for liquid xenon at 165 Kelvin.
These values are compared to the values obtained with data~\cite{lwang} and a reasonable agreement was achieved. 
With these parameters, we obtain the absolute ionization efficiency of nuclear recoils for a germanium detector and
the absolute scintillation efficiency of nuclear recoils for a liquid xenon detector at zero field as shown in Figure~\ref{fig:totaleff}. 
Utilizing Eq.(\ref{eq:mei2}) and Eq.(\ref{eq:mei3}), a dual-phase liquid xenon detector with  3 photoelectrons detected 
at an efficiency of 14\% in liquid and 50   
photonelectrons detected in gas with a multiplication factor of 24.55 and an extraction efficiency of 65\% 
would correspond to a detection threshold energy of 6.8 keV nuclear recoil energy~\cite{lwang}.

The discrimination of nuclear recoils from electronic recoils is usually performed through a well-defined nuclear recoil band and 
electronic recoil band for a given technology such as dual-phase liquid xenon detectors~\cite{xenon100,lux} and cryogenic
germanium detectors at 50 mini-Kelvin~\cite{cdms, edelweiss}. The width of bands is given by the respective energy resolution 
of nuclear recoils and electronic recoils for a given detector. All experiments treat energy resolution the same for both
nuclear recoils and electronic recoils. This is only valid if electronic noise and incomplete
charge are dominant contributions to the energy resolution while the intrinsic statistical variation is negligible for a given detector. 
However, for a dual-phase liquid xenon detector
and a germanium detector, the intrinsic statistical variation either dominates the detector energy resolution or is comparable to the 
contributions from electronic noise and incomplete charge. For example, XENON100~\cite{xe100e} reported the following energy resolution
functions: (1) $\frac{\sigma(E)}{E}$ = $(\frac{1.0}{\sqrt{E/MeV}}+11.2)\%$ for S1 only, (2)  $\frac{\sigma(E)}{E}$ = ($\frac{2.4}{\sqrt{E/MeV}}+3.5)\%$
 for S2 only,
and (3)  $\frac{\sigma(E)}{E}$ = $\frac{1.9\%}{\sqrt{E/MeV}}$for combined S1 and S2. These three cases can be described by an energy dependence of the 
resolution $\frac{\sigma(E)}{E}$ = $\frac{c_{1}}{E} + c_{2}$. It is clear that $\frac{c_{1}}{E}$ is related to the
statistical variation and $c_{2}$ describes the noise level. Note that $c_{2}$ is comparable to zero for combined S1 and S2 reported
by XENON100~\cite{xe100e}. This implies that the statistical variation dominates the energy resolution of the XENON100 detector
 for the case of combined S1 and S2. In the case of the CDMS experiment, the detector-averaged energy resolution below 10 keV was 
reported to be $\sigma(E)$ = $\sqrt{(0.016)^2+(0.0215)^2E + (0.021)^2E^2}[keV]$~\cite{cdms1, wei}. As can be seen in this equation, the term, $(0.016)^2$, is a constant from electronic noise,
 the term, $(0.0215)^2E$ represents statistical variation, and the term, $(0.021)^2E^2)$ is related to the contribution from noise and incomplete charge. The 
contributions from these three terms are comparable. 
In the case in which the statistical variation dominates the energy resolution or is comparable to the contribution from
noise and incomplete charge, the energy resolution for nuclear recoils cannot be described by the measured energy resolution
with gamma-ray sources. This is because the statistical variation of nuclear recoils events differs from that of electronic recoil events 
with the same energy. We attribute this difference to the processes of energy partition for these two classes of events.

Upon deposition of energy in a given target, $E_{0}$, two types of excitations, (a) lattice excitation with no
formation of mobile charge pairs and (b) ionization with formation of mobile charge pairs, are created by an incoming particle.  
Lattice excitations produce $N_{x}$ phonons of energy $E_{x}$. Ionization form $N_{i}$ charge pairs of energy $E_{i}$.
For an energy loss process, energy conservation requires $E_{0}$ = $E_{i}N_{i}$ + $N_{x}N_{x}$. As fluctuations in 
$N_{i}$ are compensated by fluctuations in $N_{x}$ to keep $E_{0}$ constant, 
$\frac{dE_{0}}{dN_{i}}\Delta N_{i}$ + $\frac{dE_{0}}{dN_{x}}\Delta N_{x}$ = 0 and $E_{i} \Delta N_{i}$ + $E_{x}\Delta N_{x}$ = 0. 
From averaging many events, 
one obtains for the variance: $E_{i}\sigma_{i}$ = $E_{x}\sigma_{x}$, with $\sigma_{x}$ = $\sqrt{N_{x}}$ assuming Gaussian statistics. 
Thus, $\sigma_{i}$ = $\frac{E_{x}}{E_{i}}\sqrt{N_{x}}$. Since $N_{x}$ = $\frac{E_{0}-E_{i}N_{i}}{E_{x}}$, one obtains
$\sigma_{i}$ = $\frac{E_{x}}{E_{i}}\sqrt{\frac{E_{0}}{E_{x}}-\frac{E_{i}}{E_{x}}N_{i}}$. Each ionization leads to a charge pair 
that contributes to the signal, therefore, $N_{i}$ = $\frac{E_{0}}{\varepsilon_{i}}$ and 
\begin{equation}
\sigma_{i} = \sqrt{\frac{E_{0}}{\varepsilon_{i}}}\cdot\sqrt{\frac{E_{x}}{E_{i}}(\frac{\varepsilon_{i}}{E_{i}}-1)},
\label{eq:sigma}
\end{equation}
 where $\varepsilon_{i}$ represents
mean energy expended per e-h pair. 

The statistical variation is usually quantified by the Fano factor~\cite{fano} ($F$), which is defined for any integer-valued 
random variable as the 
ratio of the variance ($\sigma_{i}^{2}$) of the variable to its mean ($N_{i}$), $F$ = $\frac{\sigma_{i}^{2}}{N_{i}}$.  
Thus, one obtains $\sigma_{i}$ = $\sqrt{FN_{i}}$. Comparing this expression to Eq. (\ref{eq:sigma}), one finds 
\begin{equation}
F = \sqrt{\frac{E_{x}}{E_{i}}(\frac{\varepsilon_{i}}{E_{i}}-1)}, 
\label{eq:fano}   
\end{equation}

\begin{figure}
\includegraphics[angle=0., width=8.cm]{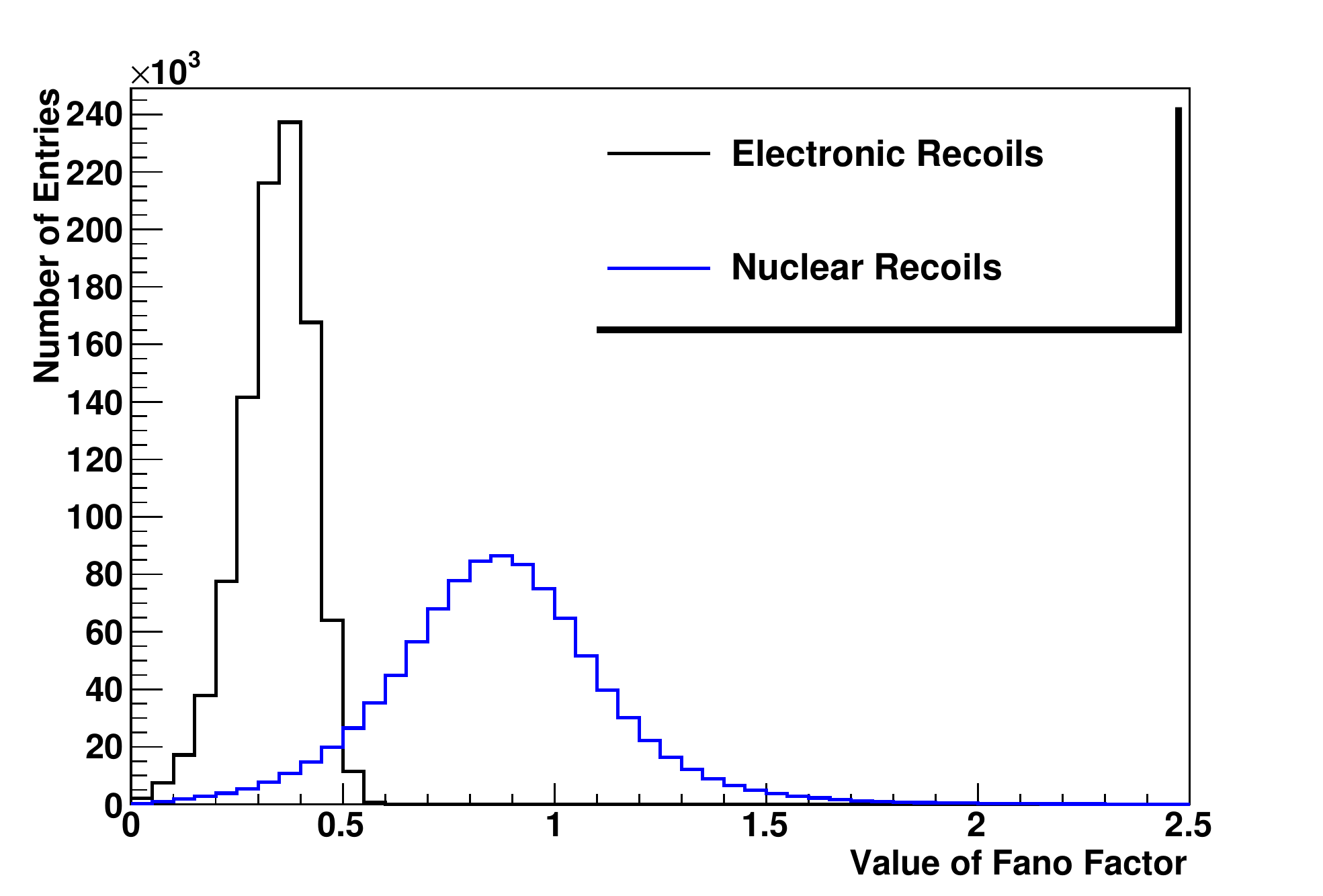}
\caption{\small {The calculated Fano factors for electronic recoils and nuclear recoils in germanium detector at 50 mini-Kelvin.}}
\label{fig:fanoge}
\end{figure}

\begin{figure}
\includegraphics [angle=0, width=8.cm]{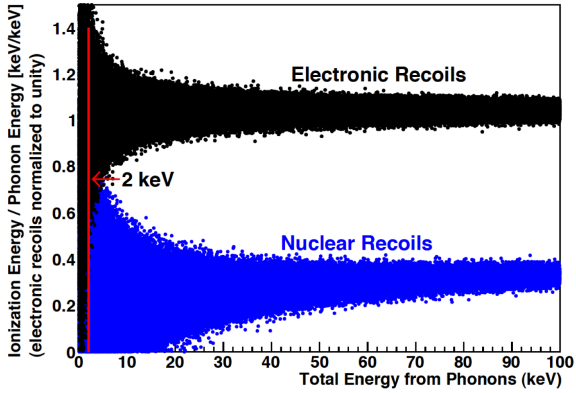}
\caption{\small {The predicted electronic recoil band and nuclear recoil band for a germanium detector at 50 mini-Kelvin. 
The red line stands for 2 keV phonon energy.}}
\label{fig:cdmsdis}
\end{figure}

In the case of the SuperCDMS-like detector, the average energy expended per e-h pair, $\varepsilon$, is critical to  
interpret the measured phonon energy and hence the measured recoil energy since $E_{r}$ = $E_{p}$ - $E_{Q}\frac{eV_{b}}{\varepsilon}$
is used to reconstruct recoil energy, where $E_{p}$ the measured phonon energy, $E_{Q}$ the ionization energy, 
and $V_{b}$ the bias voltage. However, there are several processes~\cite{follow} that reduce phonons. Therefore,  
the measured phonon energy must be corrected for efficiencies, $E_{p}$ = $\eta_{pri}(E_{r}-\frac{E_{Q}}{\varepsilon}E_{g}) +
\eta_{rec}(f_{Q}\frac{E_{Q}}{\varepsilon})E_{g}+\eta_{Luke}(f_{Q}\frac{E_{Q}}{\varepsilon}eV_{b}$), where $\eta_{pri}$, $\eta_{rec}$,
and $\eta_{Luke}$ the detection efficiencies for primary phonons, the phonons from recombination of e-h pairs, 
and Luke phonons, respectively, and
$f_{Q}$ the fraction of the total charge observed. Because of such a complicated relation of the measured phonon energy with
different efficiencies, $\varepsilon$ has to be determined independently. 
At 50 mini-Kelvin, the average primary phonon energy is about 4.141$\times$10$^{-3}$ eV with a frequency
of 1 THz~\cite{cdms2, wei}. The Neganov-Luke phonons have an average energy of 1.242$\times$10$^{-3}$ eV with a frequency of 0.3 THz~\cite{wang}. 
The band gap energy at 50 mini-Kelvin is about 0.742 eV. Using Eq.(\ref{eq:fano}) and 
the measured energy resolution related to statistical variation, 
(0.0215)$^{2}$E(keV)~\cite{wei}, one obtains a Fano factor, $F$ = 0.14 and $\varepsilon_{ER}$ = 3.32 eV
at 50 mini-Kelvin for electronic recoils.
The average energy expended per e-h pair for nuclear recoils can be expressed 
as $\varepsilon_{NR}$ = $\frac{\varepsilon_{ER}}{\varepsilon}$, where ${\varepsilon}$ is ionization efficiency, which
can be calculated using Lindhard's theory~\cite{lind} or Barker-Mei's model~\cite{barker, barker1}. Together with Eq.(\ref{eq:fano}), the 
Fano factor for nuclear recoils can be obtained as shown in Figure~\ref{fig:fanoge}. Note that the ionization yield 
for electronic recoils is normalized to be a unit. If one uses $\varepsilon_{ER}$ = 8.9 eV instead of 3 eV~\cite{cdms} in
a SuperCDMS-like detector, 
the threshold of nuclear recoil energy is determined to be $\sim$2.15 keV~\cite{wei} with 50 mini-Kelvin at 69 volts. 
As can be seen in Figure~\ref{fig:fanoge}, the Fano factor for low energy electronic recoils is a sub-Poisson distribution, while the Fano
factor for nuclear recoils is a convolution of both sub-Poisson (Fano factor $<$ 1) and super-Poisson (Fano factor $>$ 1) 
distribution depending on nuclear recoil energy. 
Consequently, the width
of nuclear recoils and the width of electronic recoils using the ratio of ionization yield to the total phonon energy are 
governed by the intrinsic statistical variation of these two classes of events as shown in Figure~\ref{fig:cdmsdis}. In the
generation of Figure~\ref{fig:cdmsdis}, the electronic recoils were randomly chosen between 0.1 keV to 100 keV, while the
nuclear recoils were simulated with $^{252}$Cf neutron energy spectrum expressed as
 $N(E_n)$ = $\sqrt{E_n}e^{(-E_n/1.565)}$~\cite{cf252}. 
It is demonstrated in Figure~\ref{fig:cdmsdis} that there is no discrimination between electronic recoils and nuclear recoils when the
phonon energy (recoil energy) is below 2 keV. The energy resolution, $\sigma(E)$, induced by this statistical variation
 for the 2 keV nuclear recoil is about 0.3 keV. Note that the $\sigma(E)$ for the 0.5 keV and 2 keV electronic recoil 
is about 0.037 and 0.085 keV, respectively. Therefore, the uncertainty on the threshold energy of nuclear recoils must 
be evaluated separately from electronic recoils. 

\begin{figure}
\includegraphics[angle=0., width=8.cm]{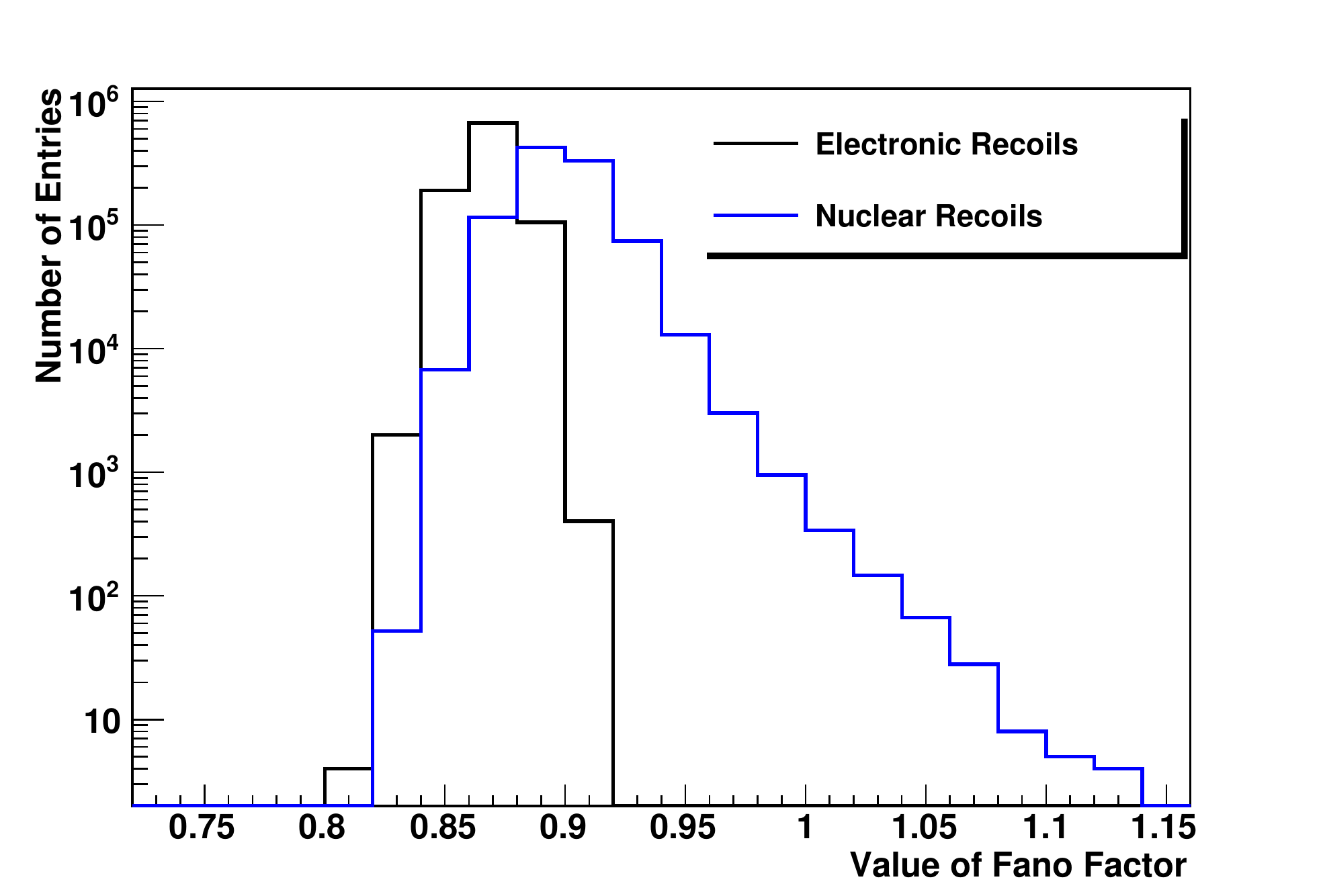}
\caption{\small {The calculated Fano factors for electronic recoils and nuclear recoils in liquid xenon detector at 165 Kelvin.}}
\label{fig:xefano}
\end{figure}

For a given dual-phase xenon detector, the Fano factor for optical photons is $F_{N}$ = $\sigma_{N}^{2}/\bar{N}$. This 
value was measured to be 0.059 by Doke et al.~\cite{doke}. Since liquid xenon detectors detect optional photons with 
photomultipliers, which convert optical photons into photoelectrons. Therefore, the Fano factor for the photoelectrons
becomes, $F_{n}$ = $\sum_{i=1}^{n} (1 + \epsilon(F_{N}-1)$~\cite{abd}, where $i$ runs from 1 to the number of photomultipliers
that are fired per event, $\epsilon$ is the
probability that an optical photon will reach the photo-detector and produce a photoelectron. 
If one assumes 
$\epsilon$ = 0.14~\cite{lux}, the Fano factor for photoelectrons is about 0.87 for electronic recoils with one photomultiplier.
 Using Eq.(\ref{eq:fano}), one
can determines $E_{x}$ = 0.047 eV if the average energy expended per electron-hole pair, $\varepsilon$ = 15.6 eV~\cite{april1}.
Thus, we can determine Fano factor for nuclear recoils using the similar method described above. Figure~\ref{fig:xefano}
shows Fano factor for electronic recoils and nuclear recoils in liquid xenon with one photomultiplier.
Depending on the
number of photomultipliers that are involved in detecting photons from an energy deposition, the Fano factor can be much larger than 1.

\begin{figure}
\includegraphics [angle=0.0, width=8.cm]{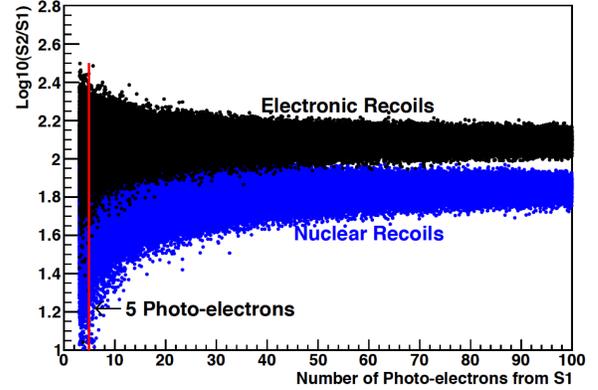}
\caption{\small {The predicted electronic recoil band and nuclear recoil band for a 
dual-phase liquid xenon detector at 165 Kelvin without considering the saturation effect in S2.
 The red line represents 5 photoelectrons, which corresponds to 
3.5 keV nuclear recoil energy.}}
\label{fig:xedis}
\end{figure}

As a result, the statistical variation dominates the widths of the nuclear recoil band and the electronic recoil band 
using the ratio of signal in gas phase (S2) to signal in liquid (S1) for
discriminating these two classes of events as shown in Figure~\ref{fig:xedis}. In the simulation of S1 and S2 signals, we
assume the product of photon detection efficiency and the photoelectron efficiency is 14\%~\cite{lux} in liquid, the 
electron extraction efficiency from liquid to gas phase is 65\%~\cite{lux}, and the photon multiplification factor of S2 is
24.55~\cite{lux}. Note that the saturation effect in S2 was not taken into account in the simulation.

As can be seen in Figure~\ref{fig:xedis}, when photoelectrons in S1 is less than 2, which represents about 1.5 keV, 
the dual-phase xenon detector loses the discrimination power. At a 3 keV nuclear recoil energy, the energy 
resolution, $\sigma(E)$, is about 0.7 keV while $\sigma(E)$ is about 0.3 keV for the 2 keV and
0.1 keV for the 0.9 keV electronic recoils. 

In conclusion, the absolute energy response functions with two parameters, $\alpha$ and $\beta$, 
describe the energy scale for electronic recoils and nuclear recoils. The values of $\alpha$ and $\beta$
can be calculated or determined using well-known gamma ray sources.  
The Fano factor for nuclear recoils dominates the energy resolution in the low energy region
for detecting WIMPs using both SuperCDMS-type and dual-phase xenon detectors. The Fano factor for nuclear recoils 
is larger than that of electronic recoils. Note that the difference in Fano factors between nuclear recoils and electronic recoils 
is mainly due to the difference in the average energy expended per e-h pair. This causes a different energy scale 
function between nuclear recoils and electronic recoils. 
Therefore, the calibration of both energy scale and energy resolution needs to
be implemented for electronic recoils and nuclear recoils separately.

The authors wish to thank Christina Keller for careful reading of this manuscript. 
This work was supported in part by NSF PHY-0919278, NSF PHY-1242640, DOE grant DE-FG02-10ER46709,
 and a governor's research center supported by the state of South Dakota. 

%
%

\end{document}